\documentstyle[12pt,overcite,procs,mystyle]{article}

\begin{document}
\pagestyle{plain}
\def\oddhead{{\rm gr-qc/9409019 \hfil UCLA/94/TEP/34}}
\thispagestyle{mystyle}
\begin{Titlepage}
\Title
ANOMALY IN 2D-GRAVITY COUPLED TO MATTER FIELDS\footnote{%
Talk given at the 7th Marcel Grossman Meeting on General Relativity,
Stanford, July 28, 1994.}
\endTitle
\Author{DANIEL CANGEMI}
Department of Physics, University of California,
Los Angeles, CA 90024-1547, USA
\endAuthor
\endAuthors

\begin{Abstract}
In 1+1 dimensions, the Wheeler-DeWitt equation cannot be imposed owing to an
anomaly in its commutator with the diffeomorphism constraint. A similar
obstruction prevents also a semiclassical definition of a WKB local time.
\end{Abstract}

\end{Titlepage}
\addtocounter{footnote}{1}

\vspace{.5in}
A consistent construction of a quantum theory of gravity is still lacking.
Superstring theory, as candidate of the theory of everything, is too premature
to show such basic phenomena as black-holes and Hawking radiation.
Semiclassical analysis tries to capture the general structure of Quantum
Gravity but encounters several difficulties like the problem of time or the
issue of information lost that raises serious concerns on the unitarity of
the full
quantum theory. It is a challenge to reconciliate these two extreme points of
view.

Quantum Gravity is usually presented in a canonical formalism where spacetime
is sliced in spacelike surfaces. The dynamics is then contained in an evolution
equation known as the Wheeler-DeWitt equation: ${\cal E} \Psi[h_{ab}, \phi] =
0$; it requires the Hamiltonian ${\cal E}$ to vanish on physical functionals
depending on the spatial components of the metric and on the matter field(s).
Invariance under diffeomorphisms of the spacelike surfaces is inforced through
an additional constraint ${\cal P} \Psi[h_{ab}, \phi] = 0$.

The semiclassical limit of the theory is achieved by a Born-Oppenheimer
expan\-sion\cite{kiefer} in the gravitational constant of the Wheeler-DeWitt
equation and leads to the Tomonaga-Schwinger functional equation: $i \hbar
\partial \chi / \partial \tau = {\cal E}_{\rm matter} \chi$, through the
introduction of a semiclassical (WKB) time functional $\tau(x; h_{ab}]$.

The subject of this talk is to address these questions in a simpler
(1+1)-dimensional context. We will show that indeed anomalies occur
in a straightforward Dirac quantization of our model;
the Wheeler-DeWitt equation and the
diffeomorphism constraint cannot be imposed simultaneously and a WKB time
cannot be introduced.

\renewcommand{\thefootnote}{\fnsymbol{footnote}}
Lineal gravity is a gauge theory\cite{quantum} based on an extension of the
Poincar\'e group. Its action\footnote{Notation: tangent
space is indexed by $(a,b,\ldots)$ with metric tensor $h_{ab} = {\rm diag}
(1,-1)$ and $\epsilon^{ab} = -\epsilon^{ba}$, $\epsilon^{01} = 1$. Also $(D_\mu
e_\nu)^a = \partial_\mu e^a_\nu + \epsilon^a{}_b \omega_\mu e^b_\nu$,
$(D_\mu q)^a = \partial_\mu q^a + \epsilon^a{}_b (\omega_\mu q^b - e^b_\mu)$,
$(D_\mu \varphi)^a = \partial_\mu \varphi^a + \epsilon^a{}_b \omega_\mu
\varphi^b$.}
\beqa
I &=& {1\over4\pi G} \int d^2x \, \epsilon^{\mu\nu} \left[ \eta_a (D_\mu
e_\nu)^a + \eta_2 \partial_\mu \omega_\nu + \eta_3 (\partial_\mu a_\nu +
\frac{1}{2} e_\mu^a \epsilon_{ab} e_\nu^b ) \right] \nonumber\\
&& + \int d^2x \, \epsilon^{\mu\nu} (D_\mu q)^a \epsilon_{ab} \biggl[ (D_\nu
\varphi)^b \varphi_3 - \varphi^b \partial_\nu \varphi_3 + (D_\nu q)^b \varphi^c
\varphi_c \biggr]
\label{action}
\eeqa
is constructed out of a {\it Zweibein} $e^a_\mu$, a spin-connection
$\omega_\mu$ and four Lagrange multipliers $\eta_a, \eta_2, \eta_3$ for the
gravity sector and out of a field multiplet $\varphi_a, \varphi_3$ and an
auxiliary field $q^a$ for the matter interaction. The auxiliary field acts as a
Higgs field to
insure gauge invariance and in the ``unitary gauge'' $q^a = 0$, this action is
(classically) equivalent to dilaton gravity coupled to massless scalar
field.\cite{CGHS}. Notice also the topological nature of the action, since it
does not depend on a background metric in the spacetime.

As in a gauge theory, quantization is straightforward. We present it in a
Schr\"odinger picture. We shift $\eta_a \to \eta_a + 2\pi G \varphi_a
\varphi_3$, $\eta_2 \to \eta_2 - 2\pi G q^a \varphi_a \varphi_3$ and eliminate
the field $\varphi_a$ so we can rewrite the Lagrangian~(\ref{action}) in a
first order formalism [time/space derivation are denoted by overdot/dash and
the index of $\varphi_3$ is omitted],
\beq
\widetilde{\cal L} = {1 \over 4\pi G}
\left( \eta_a \dot{e}_1^a
+ \eta_2 \dot{\omega}_1
+ \eta_3 \dot{a}_1 \right)
+ p_a \dot{q}^a + \Pi \dot{\varphi}
+ e_0^a G_a + \omega_0 G_2 + a_0  G_3  - u \widetilde{\cal E} - v
\widetilde{\cal P}
\label{first_order}
\eeq
where $(\eta_a, \eta_2, \eta_3) / 4\pi G$ and $(e^a_\mu, \omega_\mu, a_\mu)$
are canonically conjugate and the canonical momenta $p_a, \Pi$ are introduced
through the constraints
\beq
\widetilde{\cal E} = (D_1 q)^a \epsilon_a^{~b} p_b
+ \frac{1}{2} \left( \Pi^2 + (\varphi')^2 \right) \hspace{.5in}
\widetilde{\cal P} = p_a \left( D q \right)^a + \Pi \varphi'
\label{EP}
\eeq
The other constraints $G_a, G_2, G_3$ can be read of~(\ref{action}) and are not
given here.
They generate gauge transformation on the space of functionals $\Psi[\eta_a,
\eta_2, \eta_3, q^a, \varphi]$ and are solved by the {\it
Ansatz\/}\cite{quantum}
\beqa
\Psi[\eta_a, \eta_2, \eta_3, q^a, \varphi] &=&
\delta (\eta_3(x) - \lambda) \; e^{\frac{i}{8\pi G \lambda} \int
( \epsilon^{ab} \eta_a d\eta_b + M \, d\rho^a \epsilon_{ab} \rho^b / \rho^c
\rho_c} )\nonumber\\
&&\hspace{-1.2in} \times \int {\cal D} \gamma \;
e^{{i \over 8\pi G \lambda} \int (M - \frac{\lambda^2}{2} \rho^a \rho_a) \, d
\gamma} \;
{\Phi} (\underbrace{\sqrt{\rho^a \rho_a} \cosh \gamma}_{\equiv \, r^0},
\underbrace{\sqrt{\rho^a \rho_a} \sinh \gamma}_{\equiv \, r^1}, \varphi)
\label{psimatter}
\eeqa
with $\rho^a = q^a + \frac{1}{\lambda} \eta^a$, $M = \eta_a \eta^a - 2 \eta_2
\eta_3$.
Acting on ${\Phi}$, $\widetilde{\cal E}$ and $\widetilde{\cal P}$ become ($g =
8 \pi G / \lambda$)
\beq
{\cal E} \Phi =  \left[ - \frac{1}{2}
\left( {g \pi^a \pi_a} + {1\over g} r'^a r'_a \right)
+ \frac{1}{2} \left( \Pi^2 + \varphi'^2 \right) \right] \Phi = 0 \hspace{.4in}
{\cal P} \Phi = \left[ \pi_a r'^a + \Pi \varphi' \right] \Phi = 0
\label{EP2}
\eeq
corresponding to an effective Lagrange density
${\cal L} = \pi_a \dot{r}^a + \Pi \dot{\varphi} - u {\cal E} - v {\cal P}$.

The first equation in~(\ref{EP2}) is the Wheeler-DeWitt equation
for lineal gravity, whereas the second one reflects invariance under
diffeomorphisms. The constraints $\cal E$ and $\cal P$ have the standard form
for free massless fields $r^0, r^1, \varphi$ and usual anomalous commutation
relations.
\beqa
&& \left[ {\cal E} (x), {\cal E} (y) \right]
= \left[ {\cal P}(x), {\cal P}(y) \right]
= i \left( {\cal P} (x) + {\cal P} (y) \right) \, \delta' (x-y)
\nonumber\\[-6pt]
\label{ccr} \\[-6pt]
&& \left[ {\cal E} (x) , {\cal P} (y) \right]
= i \left( {\cal E}(x) + {\cal E}(y) \right) \,
\delta' (x-y) - i \frac{c_{r^0} + c_{r^1} + c_\varphi}{12\pi}
\delta^{\prime\prime\prime}(x - y) \nonumber
\eeqa
In the absence of matter, the wave functional has been explicitely computed
without showing any anomaly.\cite{amati} The regularization (e.g. by point
splitting) of ${\cal E}(x), {\cal P}(x)$ has thus to be done in such a way that
the gravity anomaly cancels: $c_{r^0} = - c_{r^1} = 1$. This cancellation is
possible due to opposite signs in the kinetic energy of $r^0$ and $r^1$.
However, in presence of matter fields, the anomaly ($c_{\rm tot} = c_\varphi$)
does not vanish and the constraints~(\ref{EP2}) cannot be enforced
simultaneously.

The semiclassical limit is obtained by expanding the wave functional in powers
of $g \equiv 8\pi G / \lambda$: $\Phi = \exp i\left(g^{-1}S_g + S_0 + gS_1 +
\ldots \right)$. The evaluation of the Wheeler-DeWitt equation at different
orders in $g$ gives
\beqa
&\makebox[0in][c]{$\displaystyle
g^{-2} : {\delta S_g \over \delta \varphi} = 0 \hspace{.5in}
g^{-1} : \frac{\delta S_g}{\delta r^a} \frac{\delta S_g}{\delta r_a}
+ r'^a r'_a = 0$} & \nonumber\\[-6pt]
\label{semiclassical} \\[-6pt]
& \makebox[0in][c]{$\displaystyle g^0 : \left( - {1\over2}
{\delta^2 \over \delta \varphi \delta \varphi}
+ {1\over 2}  \varphi'^{2} \right) \, e^{i S_0}
=  r'^a \epsilon_a^{~b} \, {1\over i} \, \frac{\delta}{\delta r^b} e^{i S_0}$}
& \nonumber
\eeqa
The first equation shows that $S_g$ corresponds to the phase for vanishing
matter fields and the second equation identifies $S_g$ with the phase of pure
gravity found in Ref.~[\citen{amati}]. The third equation is the usual
Tomonaga-Schwinger equation once the right side is defined to be a functional
derivative with respect to some local time coordinate $\tau(x; r^a]$. However
such a definition is impossible since $\delta / \delta \tau(x; r^a]$ commutes
with itself at all points on the line, but $r'^a(x) \epsilon_a^{~b} {\delta
\over \delta r^b (x)}$ does not.

The quantum constraints of lineal gravity with matter are anomalous.
It could be that this anomaly may be suppressed by a different normal
ordering\cite{kuchar} in the quantum operators. Alternatively, a part of the
constraints must be relaxed or the initial model must be modified in order to
define an acceptable lineal
quantum gravity. Whether similar problems arise in 3+1 dimensions is an open
question, and an understanding of the lower dimensional theory is certainly
necessary.

This work is supported by the U.S. National Science Foundation under contract
\#PHY-92-18990 and by the Swiss National Science Foundation. I thank R.~Jackiw
for his collaboration in this study and A.~Mikovic for an interesting comment.

\end{document}